\documentclass[secnumarabic, prb]{revtex4}
\usepackage{bm}

\begin{document}

\title{Relativistic force transformation}
\author{Valery P.\ Dmitriyev}
\affiliation{Lomonosov University, P.O.\ Box 160, Moscow, 117574
Russia}


\begin{abstract}
Formulae relating one and the same force  in two inertial frames of reference are derived directly from the
Lorentz transformation of space and time coordinates and relativistic equation for the dynamic law of motion in
three dimensions. We obtain firstly relativistic transformation for the velocity and acceleration of a particle.
Then we substitute them in the relativistic dynamic equation and perform tedious algebraic manipulations. No
recourse were made to "general rules for the transformation of 4-tensors". Formulae obtained were verified in
electrodynamics.
\end{abstract}


\maketitle

\section {Introduction}
The relativistic mechanics looks in one dimension
as\cite{Goldstein}
\begin{equation}
\frac{\ddot{x}}{(1-\dot{x}^2/c^2)^{3/2}}=F\label{1}
\end{equation}
where $\dot{x}=dx/dt$, $\ddot{x}=d^2x/dt^2$. Equation (\ref{1}) is
invariant under the Lorentz transformation
\begin{eqnarray}
x'&=&\frac{x-vt}{(1-v^2/c^2)^{1/2}}, \label{2}\\
y'&=& y, \quad z'=z,\label{3}\\
t'&=&\frac{t-xv/c^2}{(1-v^2/c^2)^{1/2}}\label{4}
\end{eqnarray}
where $v$ is the parameter that has the meaning of the velocity
\begin{equation}
\textbf{v} = (v,0,0) \label{5}
\end{equation}
which the inertial frame of reference $K'$ moves in the inertial
frame of reference $K$.  Finding from (\ref{2}), (\ref{4})
relativistic transformations of the velocity $\dot{x}$ and
acceleration $\ddot{x}$ of the body and substituting them in
\begin{equation}
\frac{\ddot{x}'}{(1-\dot{x}'^2/c^2)^{3/2}}=F'\label{6}
\end{equation}
we may verify that the left-hand part of (\ref{6}) turns exactly
into the left-hand part of (\ref{1}). Hence, we have for the
right-hand parts of equations (\ref{1}) and (\ref{6})
\begin{equation}
F' = F. \label{7}
\end{equation}

In three dimensions the situation is complicated. The left-hand
parts of scalar dynamic equations in $K'$ are expressed as linear
combinations of their left-hand parts in $K$. This induces
respective transformation of the force $\textbf{F}$. To find
linear relations connecting each of $F'_x$, $F'_y$ and $F'_y$ with
$F_x$, $F_y$ and $F_z$ is the aim of the present work. We will
proceed in the following succession.

Firstly, the one-dimensional Lorentz transformation
(\ref{2})-(\ref{4}) will be generalized to three dimensions. Then
we will find from it the relativistic transformations of the
velocity $\dot{\textbf{r}}= d\textbf{r}/dt$ and acceleration
$\ddot{\textbf{r}}= d^2\textbf{r}/dt^2$, where $\textbf{r}
=(x,y,z)$. We will substitute them into a three-dimensional
generalization of the dynamic equation (\ref{1}) and after tedious
manipulations find the relativistic transformation of
$\textbf{F}$. Finally, we will apply the result to the system of
two electric charges moving with a constant velocity.

\section{Three dimensional Lorentz transformation}

Let $\textbf{v}$ be arbitrarily oriented in space. We have from
(\ref{2}) and (\ref{4}) for the projection of $\textbf{r}$  on the
direction of $\textbf{v}$
\begin{eqnarray}
\textbf{r}'\cdot\textbf{v}/v &=&
\gamma(\textbf{r}\cdot\textbf{v}/v-vt),
\label{8}\\
t' &=& \gamma(t - \textbf{r}\cdot\textbf{v}/c^2)\label{9}
\end{eqnarray}
where
\begin{equation}
\gamma = \frac{1}{(1-v^2/c^2)^{1/2}}. \label{10}
\end{equation}
By (\ref{3}) the direction perpendicular to $\textbf{v}$ remains
unchanged:
\begin{equation}
\textbf{r}'_\bot= \textbf{r}_\bot. \label{11}
\end{equation}
Expanding a vector into the sum of vectors perpendicular and
parallel to $\textbf{v}$ we get
\begin{equation}
\textbf{r}= \textbf{r}_\bot+ \textbf{r}_\|. \label{12}
\end{equation}
This gives, using (\ref{8}), (\ref{11}) and (\ref{12})
\begin{equation}
\textbf{r}'= \textbf{r}'_\bot+ \textbf{r}'_\| = \textbf{r}'_\bot
+(\textbf{r}'\cdot\textbf{v}/v)\textbf{v}/v \\
= \textbf{r}_\bot +
\gamma(\textbf{r}\cdot\textbf{v}/v-vt)\textbf{v}/v = \textbf{r} +
(\gamma -1)\textbf{r}_\| - \gamma \textbf{v}t. \label{13}
\end{equation}

\section{Transformation of velocity}

We have from (\ref{9}) and (\ref{13})
\begin{eqnarray}
dt' &=& \gamma(dt-d\textbf{r}\cdot\textbf{v}/c^2),\label{14}\\
d\textbf{r}' &=& d\textbf{r} + (\gamma -1)d\textbf{r}_\| - \gamma
\textbf{v}dt \label{15}.
\end{eqnarray}
We find from (\ref{14}) and (\ref{15})
\begin{equation}
\dot{\textbf{r}}' = \frac{d\textbf{r}'}{dt'} = \frac{\dot{\textbf{r}}+(\gamma-1)\dot{\textbf{r}}_\| - \gamma
\textbf{v}}{\gamma (1-\dot{\textbf{r}}\cdot\textbf{v}/c^2)} =
\frac{\dot{\textbf{r}}+\textbf{v}[(\gamma-1)\dot{\textbf{r}}\cdot\textbf{v}/v^2 - \gamma ]}{\gamma
(1-\dot{\textbf{r}}\cdot\textbf{v}/c^2)} . \label{16}
\end{equation}
If $\textbf{v}$ is directed along the $x$-axis then we may get
from (\ref{16}) and (\ref{5})
\begin{eqnarray}
\dot{x}' &=& \frac{\dot{x}-v}{1-\dot{x}v/c^2}, \label{17}\\
\dot{y}' &=& \frac{\dot{y}}{\gamma(1-\dot{x}v/c^2)},\quad \dot{z}'
= \frac{\dot{z}}{\gamma(1-\dot{x}v/c^2)}. \label{18}
\end{eqnarray}
The following useful relation can be obtained from (\ref{17}) and (\ref{18})
\begin{equation}
\frac{1}{(1-\dot{\textbf{r}}'^2/c^2)^{1/2}} =
\frac{\gamma(1-\dot{x}v/c^2)}{(1-\dot{\textbf{r}}^2/c^2)^{1/2}}
\label{19}
\end{equation}
where
\begin{equation}
\dot{\textbf{r}}^2=\dot{x}^2 + \dot{y}^2 + \dot{z}^2. \label{20}
\end{equation}

\section{Transformation of acceleration}

We have from (\ref{14}) for the case of (\ref{5})
\begin{equation}
dt' = \gamma(dt-dx v/c^2) = dt\gamma(1- \dot{x} v/c^2).\label{21}
\end{equation}
Differentiating (\ref{17}) and using it and (\ref{21}) we get
\begin{equation}
\ddot{x}' = \frac{d\dot{x}'}{dt'} =
\frac{d\dot{x}'}{dt}\frac{dt}{dt'} = \left[\frac{\ddot{x}}{1-
\dot{x} v/c^2}+\frac{(\dot{x}-v)\ddot{x}v/c^2}{(1- \dot{x}
v/c^2)^2}\right]\frac{1}{\gamma(1- \dot{x} v/c^2)}.\label{22}
\end{equation}
Using (\ref{10}) in (\ref{22}) gives finally
\begin{equation}
\ddot{x}'  = \frac{\ddot{x}}{[\gamma(1- \dot{x}
v/c^2)]^3}.\label{23}
\end{equation}
Differentiating (\ref{18}) and using it and (\ref{21}) we get for
a transverse acceleration
\begin{equation}
\ddot{y}' = \frac{d\dot{y}'}{dt'} =
\frac{d\dot{y}'}{dt}\frac{dt}{dt'} =
\gamma^{-1}\left[\frac{\ddot{y}}{1- \dot{x}
v/c^2}+\frac{\dot{y}\ddot{x}v/c^2}{(1- \dot{x}
v/c^2)^2}\right]\frac{1}{\gamma(1- \dot{x} v/c^2)}.\label{24}
\end{equation}
Relation (\ref{24}) gives
\begin{equation}
\ddot{y}' =
\frac{1}{\gamma^2(1-\dot{x}v/c^2)^2}(\ddot{y}+\ddot{x}\frac{\dot{y}v/c^2}{1-\dot{x}v/c^2}).\label{25}
\end{equation}
The analogous expression for $z$ is
\begin{equation}
\ddot{z}' =
\frac{1}{\gamma^2(1-\dot{x}v/c^2)^2}(\ddot{z}+\ddot{x}\frac{\dot{z}v/c^2}{1-\dot{x}v/c^2}).\label{26}
\end{equation}

\section{Transformation of $\textbf{F}_\|$}

The three-dimensional relativistic mechanics is\cite{Goldstein}
\begin{equation}
\frac{d}{dt}\left[\frac{m\dot{\textbf{r}}}{(1-\dot{\textbf{r}}^2/c^2)^{1/2}}\right]=\textbf{F}.\label{27}
\end{equation}
Fulfilling the differentiation in (\ref{27}) and taking scalar
components
\begin{eqnarray}
\frac{m\ddot{x}}{(1-\dot{\textbf{r}}^2/c^2)^{1/2}}+\frac{m\dot{x}(\dot{\textbf{r}}\cdot\ddot{\textbf{r}})/c^2}{(1-\dot{\textbf{r}}^2/c^2)^{3/2}}=F_x,\label{28}\\
\frac{m\ddot{y}}{(1-\dot{\textbf{r}}^2/c^2)^{1/2}}+\frac{m\dot{y}(\dot{\textbf{r}}\cdot\ddot{\textbf{r}})/c^2}{(1-\dot{\textbf{r}}^2/c^2)^{3/2}}=F_y,\label{29}\\
\frac{m\ddot{z}}{(1-\dot{\textbf{r}}^2/c^2)^{1/2}}+\frac{m\dot{z}(\dot{\textbf{r}}\cdot\ddot{\textbf{r}})/c^2}{(1-\dot{\textbf{r}}^2/c^2)^{3/2}}=F_z.\label{30}
\end{eqnarray}
Strictly speaking, equation (\ref{27}) is not Lorentz invariant.
However, we may retain the form of (\ref{28}) in $K'$ system:
\begin{equation}
\frac{m\ddot{x}'}{(1-\dot{\textbf{r}}'^2/c^2)^{1/2}}+\frac{m\dot{x}'(\dot{\textbf{r}}'\cdot\ddot{\textbf{r}}')/c^2}{(1-\dot{\textbf{r}}'^2/c^2)^{3/2}}=F'_x.\label{31}
\end{equation}
Substituting (\ref{17}), (\ref{18}), (\ref{23}), (\ref{25}) and
(\ref{26}) in (\ref{31}), the left-hand part of (\ref{31}) can be
represented as a linear combination of left-hand parts of
equations (\ref{28}), (\ref{29}) and (\ref{30}). This means that
retaining the form of (\ref{27}) we must transform the right-hand
part of (\ref{27}). The component $F'_x$ of the force is
represented as respective linear combination of $F_x$, $F_y$ and
$F_z$. Next, we will perform explicitly the procedure mentioned.

Using (\ref{19}) and (\ref{23}) in (\ref{31}) gives
\begin{equation}
F'_x = \frac{m\ddot{x}}{\gamma^2(1-\dot{x}v/c^2)^2
(1-\dot{\textbf{r}}^2/c^2)^{1/2}}+\frac{m\dot{x}'(\dot{x}'\ddot{x}'+\dot{y}'\ddot{y}'+\dot{z}'\ddot{z}')\gamma^3(1-\dot{x}v/c^2)^3}{c^2(1-\dot{\textbf{r}}^2/c^2)^{3/2}}.\label{32}
\end{equation}
Then, substituting (\ref{23}), (\ref{25}), (\ref{26}) and
(\ref{17})-(\ref{18}) in the second term of (\ref{32})
\begin{eqnarray}
F'_x &=& \frac{m\ddot{x}}{\gamma^2(1-\dot{x}v/c^2)^2
(1-\dot{\textbf{r}}^2/c^2)^{1/2}}\nonumber \\
&+&\frac{m(\dot{x}-v)}{c^2(1-\dot{\textbf{r}}^2/c^2)^{3/2}} \left[
\frac{(\dot{x}-v)\ddot{x}}{(1-\dot{x}v/c^2)^2}
+\frac{\dot{y}}{1-\dot{x}v/c^2}(\ddot{y}+\ddot{x}\frac{\dot{y}v/c^2}{1-\dot{x}v/c^2})
+
\frac{\dot{z}}{1-\dot{x}v/c^2}(\ddot{z}+\ddot{x}\frac{\dot{z}v/c^2}{1-\dot{x}v/c^2})
\right] \nonumber \\
&=& \frac{m\ddot{x}}{\gamma^2(1-\dot{x}v/c^2)^2
(1-\dot{\textbf{r}}^2/c^2)^{1/2}}\nonumber \\
&+&
\frac{m(\dot{x}-v)}{c^2(1-\dot{\textbf{r}}^2/c^2)^{3/2}(1-\dot{x}v/c^2)}
\left[ \ddot{x}\frac{\dot{x}-v}{1-\dot{x}v/c^2}
+\dot{y}\ddot{y}+\ddot{x}\frac{\dot{y}^2v/c^2}{1-\dot{x}v/c^2}
+\dot{z}\ddot{z}+\ddot{x}\frac{\dot{z}^2v/c^2}{1-\dot{x}v/c^2}
\right] .\label{33}
\end{eqnarray}
Firstly, we consider the portion of (\ref{33}) that contains
$\ddot{x}$. Using in it (\ref{10}) and (\ref{20}) gives
\begin{eqnarray}
\frac{m\ddot{x}}{(1-\dot{\textbf{r}}^2/c^2)^{1/2}(1-\dot{x}v/c^2)^2}
\left[ 1-v^2/c^2 +
\frac{(\dot{x}-v)^2+(\dot{y}^2+\dot{z}^2)(\dot{x}-v)v/c^2}{c^2(1-\dot{\textbf{r}}^2/c^2)}
\right] .\label{34}
\end{eqnarray}
The expression in quadratic brackets of (\ref{34})
\begin{eqnarray}
&& 1-v^2/c^2 + \frac{(\dot{x}-v)^2 -
(1-\dot{\textbf{r}}^2/c^2)(\dot{x}-v)v + (\dot{x}-v)v
-\dot{x}^2(\dot{x}-v)v/c^2}{c^2(1-\dot{\textbf{r}}^2/c^2)}\nonumber\\
&=&1-\dot{x}v/c^2+\frac{\dot{x}^2 - \dot{x}v
-\dot{x}^2(\dot{x}-v)v/c^2}{c^2(1-\dot{\textbf{r}}^2/c^2)}=1-\dot{x}v/c^2+\frac{
\dot{x}(\dot{x}-v)(1-\dot{x}v/c^2)}{c^2(1-\dot{\textbf{r}}^2/c^2)}.
\label{35}
\end{eqnarray}
Substituting (\ref{35}) in (\ref{34})
\begin{eqnarray}
&&\frac{m\ddot{x}}{(1-\dot{\textbf{r}}^2/c^2)^{1/2}(1-\dot{x}v/c^2)}
\left[ 1 +
\frac{\dot{x}(\dot{x}-v)}{c^2(1-\dot{\textbf{r}}^2/c^2)}
\right]\nonumber\\
&=&\frac{m\ddot{x}}{(1-\dot{\textbf{r}}^2/c^2)^{1/2}}+\frac{m\ddot{x}}{(1-\dot{\textbf{r}}^2/c^2)^{1/2}(1-\dot{x}v/c^2)}
\left[ \frac{\dot{x}v}{c^2} +
\frac{\dot{x}(\dot{x}-v)}{c^2(1-\dot{\textbf{r}}^2/c^2)} \right]\nonumber\\
&=&\frac{m\ddot{x}}{(1-\dot{\textbf{r}}^2/c^2)^{1/2}}+\frac{m\ddot{x}\dot{x}(\dot{x}-\dot{\textbf{r}}^2v/c^2)}{c^2(1-\dot{\textbf{r}}^2/c^2)^{3/2}(1-\dot{x}v/c^2)}
 .\label{36}
\end{eqnarray}
We have for members from (\ref{33}) containing $\ddot{y}$ and
$\ddot{z}$
\begin{eqnarray}
\frac{m(\dot{x}-v)}{c^2(1-\dot{\textbf{r}}^2/c^2)^{3/2}(1-\dot{x}v/c^2)}
(\dot{y}\ddot{y}+\dot{z}\ddot{z}).\label{37}
\end{eqnarray}
Summing (\ref{36}) and (\ref{37}) and using
$\dot{x}\ddot{x}+\dot{y}\ddot{y}+\dot{z}\ddot{z}=\dot{\textbf{r}}\cdot\ddot{\textbf{r}}$
\begin{eqnarray}
F'_x&=&\frac{m\ddot{x}}{(1-\dot{\textbf{r}}^2/c^2)^{1/2}}
+\frac{m[\ddot{x}\dot{x}(\dot{x}-\dot{\textbf{r}}^2v/c^2)+(\dot{x}-v)(\dot{y}\ddot{y}+\dot{z}\ddot{z})]}{c^2(1-\dot{\textbf{r}}^2/c^2)^{3/2}(1-\dot{x}v/c^2)}\nonumber
\\
&=&\frac{m\ddot{x}}{(1-\dot{\textbf{r}}^2/c^2)^{1/2}}
+\frac{m[\dot{x}(\dot{\textbf{r}}\cdot\ddot{\textbf{r}})-v(\dot{x}\ddot{x}\dot{\textbf{r}}^2/c^2+\dot{y}\ddot{y}+\dot{z}\ddot{z})]}{c^2(1-\dot{\textbf{r}}^2/c^2)^{3/2}(1-\dot{x}v/c^2)}\nonumber\\
&=&\frac{m\ddot{x}}{(1-\dot{\textbf{r}}^2/c^2)^{1/2}}
+\frac{m[\dot{x}(\dot{\textbf{r}}\cdot\ddot{\textbf{r}})-v(\dot{\textbf{r}}\cdot\ddot{\textbf{r}})
+v\dot{x}\ddot{x}(1-\dot{\textbf{r}}^2/c^2)]}{c^2(1-\dot{\textbf{r}}^2/c^2)^{3/2}(1-\dot{x}v/c^2)}
 .\label{38}
\end{eqnarray}
Using (\ref{28}) in (\ref{38})
\begin{eqnarray}
F'_x&=&F_x
+\frac{m[\dot{x}(\dot{\textbf{r}}\cdot\ddot{\textbf{r}})\dot{x}v/c^2-v(\dot{\textbf{r}}\cdot\ddot{\textbf{r}})
+v\dot{x}\ddot{x}(1-\dot{\textbf{r}}^2/c^2)]}{c^2(1-\dot{\textbf{r}}^2/c^2)^{3/2}(1-\dot{x}v/c^2)}\nonumber\\
&=&F_x
+\frac{mv[-(\dot{\textbf{r}}\cdot\ddot{\textbf{r}})(1-\dot{x}^2/c^2)+\dot{x}\ddot{x}(1-\dot{\textbf{r}}^2/c^2)]}{c^2(1-\dot{\textbf{r}}^2/c^2)^{3/2}(1-\dot{x}v/c^2)}.\label{39}
\end{eqnarray}
Using (\ref{20}) in (\ref{39})
\begin{eqnarray}
F'_x &=&F_x
-\frac{mv[(\dot{\textbf{r}}\cdot\ddot{\textbf{r}})(\dot{y}^2+\dot{z}^2)/c^2+(\dot{y}\ddot{y}+\dot{z}\ddot{z})(1-\dot{\textbf{r}}^2/c^2)]}{c^2(1-\dot{\textbf{r}}^2/c^2)^{3/2}(1-\dot{x}v/c^2)}\nonumber\\
&=&F_x
-\frac{v/c^2}{(1-\dot{x}v/c^2)}\left\{\left[\frac{m\ddot{y}}{(1-\dot{\textbf{r}}^2/c^2)^{1/2}}+\frac{m\dot{y}(\dot{\textbf{r}}\cdot\ddot{\textbf{r}})/c^2}{(1-\dot{\textbf{r}}^2/c^2)^{3/2}}\right]\dot{y}
+\left[\frac{m\ddot{z}}{(1-\dot{\textbf{r}}^2/c^2)^{1/2}}+\frac{m\dot{z}(\dot{\textbf{r}}\cdot\ddot{\textbf{r}})/c^2}{(1-\dot{\textbf{r}}^2/c^2)^{3/2}}\right]\dot{z}\right\}
 .\label{40}
\end{eqnarray}
Using (\ref{29}) and (\ref{30}) in (\ref{40}) we get finally
\begin{equation}
F'_x =F_x -(F_y\dot{y}+F_z\dot{z})\frac{v/c^2}{(1-\dot{x}v/c^2)}.\label{41}
\end{equation}

\section{Transformation of $F_\bot$}

Using (\ref{18}), (\ref{19}), (\ref{25}) and  (\ref{17}),
(\ref{23}), (\ref{26}) in
\begin{equation}
\frac{m\ddot{y}'}{(1-\dot{\textbf{r}}'^2/c^2)^{1/2}}+
\frac{m\dot{y}'(\dot{\textbf{r}}'\cdot\ddot{\textbf{r}}')/c^2}{(1-\dot{\textbf{r}}'^2/c^2)^{3/2}}=F'_y\label{42}
\end{equation}
we obtain
\begin{eqnarray}
F'_y &=& \frac{m}{\gamma (1-\dot{x}v/c^2)}
\large{\{}\frac{\ddot{y}}{(1-\dot{\textbf{r}}^2/c^2)^{1/2}}+\frac{\dot{y}\ddot{x}v/c^2}{(1-\dot{\textbf{r}}^2/c^2)^{1/2}(1-\dot{x}v/c^2)}
\nonumber\\&+&
\frac{\dot{y}}{c^2(1-\dot{\textbf{r}}^2/c^2)^{3/2}}\left[\frac{(\dot{x}-v)\ddot{x}}{1-\dot{x}v/c^2}+\dot{y}\ddot{y}
+ \frac{\ddot{x}\dot{y}^2v/c^2}{1-\dot{x}v/c^2}+\dot{z}\ddot{z} +
\frac{\ddot{x}\dot{z}^2v/c^2}{1-\dot{x}v/c^2}\right]\}\nonumber\\
&=&\frac{m}{\gamma (1-\dot{x}v/c^2)} \{\frac{\ddot{y}}{(1-\dot{\textbf{r}}^2/c^2)^{1/2}} \nonumber\\&+&
\frac{\dot{y}}{c^2(1-\dot{\textbf{r}}^2/c^2)^{3/2}}\left[\frac{(1-\dot{\textbf{r}}^2/c^2)\ddot{x}v+(\dot{x}-v)\ddot{x}+(\dot{y}^2+\dot{z}^2)\ddot{x}v/c^2}{1-\dot{x}v/c^2}+\dot{y}\ddot{y}
+ +\dot{z}\ddot{z}
\right]\}\nonumber\\
&=&\frac{m}{\gamma (1-\dot{x}v/c^2)} \left\{\frac{\ddot{y}}{(1-\dot{\textbf{r}}^2/c^2)^{1/2}} +
\frac{\dot{y}}{c^2(1-\dot{\textbf{r}}^2/c^2)^{3/2}}\left[\frac{(-\dot{x}^2/c^2)\ddot{x}v+\dot{x}\ddot{x}}{1-\dot{x}v/c^2}+\dot{y}\ddot{y}
+ +\dot{z}\ddot{z}
\right]\right\}\nonumber\\
&=&\frac{m}{\gamma (1-\dot{x}v/c^2)} \left\{\frac{\ddot{y}}{(1-\dot{\textbf{r}}^2/c^2)^{1/2}} +
\frac{\dot{y}(\dot{\textbf{r}}\cdot\ddot{\textbf{r}})/c^2}{(1-\dot{\textbf{r}}^2/c^2)^{3/2}}\right\}
 .\label{43}
\end{eqnarray}
Comparing (\ref{43}) with (\ref{29}) and using in it (\ref{10}) we
get finally
\begin{equation}
F'_y = F_y\frac{(1-v^2/c^2)^{1/2}}{1-\dot{x}v/c^2}.\label{44}
\end{equation}
Similarly for $z$ component
\begin{equation}
F'_z = F_z\frac{(1-v^2/c^2)^{1/2}}{1-\dot{x}v/c^2}.\label{45}
\end{equation}

\section{Relativistic electrodynamics}

Let two particles at $(0,0,0)$ and $(x,y,z)$ be at rest in the
reference system $K$. They interact with a force
$\breve{\textbf{F}}$ that can be calculated from some field
equations. Next, let these particles move with a constant velocity
\begin{equation}
\dot{\textbf{r}} = (\dot{x},0,0).\label{46}
\end{equation}
We may calculate the force $\textbf{F}$ acted between moving
particles from the same field equations. A force can be expanded
into the sum of longitudinal and transverse components
\begin{equation}
\textbf{F} = \textbf{F}_\| + \textbf{F}_\bot.\label{47}
\end{equation}
Let us pass to the reference system $K'$ given by (\ref{5}) with
\begin{equation}
v = \dot{x}.\label{48}
\end{equation}
Then, according to (\ref{41}) with (\ref{46})
\begin{equation}
\textbf{F}'_\| = \textbf{F}_\|,\label{49}
\end{equation}
according to (\ref{45}) with (\ref{48}) and (\ref{10})
\begin{equation}
\textbf{F}'_\bot = \gamma\textbf{F}_\bot.\label{50}
\end{equation}
By (\ref{47}), (\ref{49}) and (\ref{50})
\begin{equation}
\textbf{F}' = \textbf{F}'_\|+\textbf{F}'_\bot=\textbf{F}_\|+\gamma\textbf{F}_\bot.\label{51}
\end{equation}
The principle of relativity states that we must have
\begin{equation}
\textbf{F}' = \breve{\textbf{F}}\label{52}
\end{equation}
 when $x'=x$, $y'=y$ and $z'=z$. Further we will verify (\ref{52}) for the case of two
electric charges.

We have for two charges $q_1$ and $q_2$ at rest
\begin{equation}
\breve{\textbf{F}} =
q_1q_2\frac{x\textbf{i}_x+y\textbf{i}_y+z\textbf{i}_z}{(x^2+y^2+z^2)^{3/2}}.\label{53}
\end{equation}
When a charge $q_1$ moves with a constant velocity $\dot{x}$ we must solve equations
\begin{eqnarray}
\frac{\partial^2\varphi}{\partial x^2}+\frac{\partial^2\varphi}{\partial y^2}+\frac{\partial^2\varphi}{\partial
z^2}-\frac{1}{c^2}\frac{\partial^2\varphi}{\partial t^2} &=& -4\pi q_1\delta(x-\dot{x}t,y,z),\label{54}\\
\frac{\partial^2 A_x}{\partial x^2}+\frac{\partial^2 A_x}{\partial
y^2}+\frac{\partial^2 A_x}{\partial
z^2}-\frac{1}{c^2}\frac{\partial^2 A_x}{\partial t^2} &=&
-\frac{4\pi\dot{x}}{c} q_1\delta(x-\dot{x}t,y,z).\label{55}
\end{eqnarray}
Using the Lorentz transform (\ref{2})-(\ref{4}) with (\ref{48}) we
may pass in (\ref{54}) and (\ref{55}) to reference system $K'$.
The left-hand parts of equations (\ref{54}) and (\ref{55}) are
known to be Lorentz-invariant. In $K'$ the charge is at rest,
hence fields $\varphi$ and $\textbf{A}$ do not depend on $t'$.
Using the property of $\delta$-function
$\delta(|a|x)=\delta(x)/|a|$ we obtain from (\ref{54}) and
(\ref{55})
\begin{eqnarray}
\frac{\partial^2\varphi}{\partial x'^2}+\frac{\partial^2\varphi}{\partial y^2}+\frac{\partial^2\varphi}{\partial
z^2} &=& -4\pi q_1\gamma\delta(x',y,z),\label{56}\\
\frac{\partial^2 A_x}{\partial x'^2}+\frac{\partial^2
A_x}{\partial y^2}+\frac{\partial^2 A_x}{\partial z^2} &=& -4\pi
q_1\gamma\frac{\dot{x}}{c}\delta(x',y,z).\label{57}
\end{eqnarray}
Solving equations (\ref{56}) and (\ref{57}) we get with
(\ref{48}), (\ref{2}) and (\ref{10})
\begin{eqnarray}
\varphi &=& \gamma \frac{q_1}{R},\label{58}\\
A_x&=&\gamma \frac{v}{c}\frac{q_1}{R},\quad A_y=0,\quad
A_z=0,\label{59}\\
R &=& [\gamma(x-vt)^2+y^2+z^2]^{1/2}.\label{60}
\end{eqnarray}
Calculating from (\ref{58})-(\ref{60}) the Lorentz force that acts
on a charge $q_2$ which moves in $K$ with the same velocity
$\textbf{v}$ we may obtain\cite{Dmitriyev}
\begin{eqnarray}
\textbf{F} &=& q_2[-\bm{\nabla}\varphi-\frac{1}{c}\frac{\partial\textbf{A}}{\partial t}+\frac{1}{c}(\textbf{v}\times \texttt{curl}\textbf{A})]\nonumber\\
&=&
q_1q_2\frac{\gamma(x-vt)\textbf{i}_x+\gamma^{-1}(y\textbf{i}_y+z\textbf{i}_z)}{[\gamma^2(x-vt)^2+y^2+z^2]^{3/2}}.\label{61}
\end{eqnarray}
We may isolate in (\ref{61}) longitudinal
$\textbf{F}_\|=F_x\textbf{i}_x$ and transverse
$\textbf{F}_\bot=F_y\textbf{i}_y+F_z\textbf{i}_z$ components
according to (\ref{47}), then substitute them into (\ref{51}) and
use (\ref{2}) with (\ref{10}) in the result. This gives
\begin{equation}
\textbf{F}' =
q_1q_2\frac{x'\textbf{i}_x+y\textbf{i}_y+z\textbf{i}_z}{(x'^2+y^2+z^2)^{3/2}}.\label{62}
\end{equation}
Comparing (\ref{62}) and (\ref{53}) for $x=x'$ we
confirm\cite{Dmitriyev} formula (\ref{52}).


\end{document}